\documentclass{aa}

\usepackage[colorlinks,citecolor=blue]{hyperref}
\usepackage{enumerate}
\usepackage{epsfig}
\usepackage{color}
\usepackage{txfonts}
\usepackage{natbib}
\usepackage{multirow}
\usepackage{ulem}
\usepackage{media9}
\usepackage{hyperref}

\newcommand{\Msun}{M\ensuremath{_\odot}\,}

\newcommand{\Msunpcc}{M\ensuremath{_\odot}~pc\ensuremath{^{-3} }\,}

\newcommand{\Oo}{\displaystyle}

\newcommand{\paptwo}{Mastrobuono-Battisti et al., submitted.}

\newcommand{\mywebpage}{\href{http://sergeykhoperskov.com/globular.clusters.xhtml}{http://sergeykhoperskov.com/globular.clusters}}

\begin{document}

\title{Mergers, tidal interactions, and mass exchange\\ in a population of disc globular clusters}

\titlerunning{Mergers among disc globular clusters}

\author{Sergey Khoperskov$^1$, Alessandra Mastrobuono-Battisti$^2$, Paola Di Matteo$^1$, Misha Haywood$^1$}
\authorrunning{Khoperskov et al}

\institute{GEPI, Observatoire de Paris, PSL Research University, CNRS, Place Jules Janssen, F-92195 Meudon Cedex, France \and Max Planck Institute f\"ur Astronomie, Konigsthul 17, D69117, Heidelberg, Germany}

%\date{Received ; accepted }

\abstract{We present the results of a  self-consistent $N$-body simulation following the evolution of a primordial population of thick disc globular clusters~(GCs). We study how the internal properties of such clusters evolve under the action of mutual interactions, while they orbit a Milky Way-like galaxy. For the first time, through analytical and numerical considerations, we find that physical encounters between disc GCs are a crucial factor that contributed to the shape of the current properties of the Galactic GC system. Close passages or motion on similar orbits may indeed have a significant impact on the internal structure of clusters, producing multiple gravitationally bound sub-populations through the exchange of mass and even mergers. Our model produces two major mergers and a few small mass exchanges between pairs of GCs. Two of our GCs accrete stars from two companions, ending up with three internal sub-populations. We propose these early interactions and mergers between thick disc GCs with slightly different initial chemical compositions as a possible explanation for the presence of the spreads in metallicity observed in some of the massive Milky Ways GCs.}

\maketitle

\section{Introduction}

With ages typically greater than $10$~Gyr \citep{marinfranch09, carretta10, vandenberg13}, globular clusters~(hereafter GCs)  are the oldest stellar systems in our Galaxy, and, as such, they are tracers of its early formation and mass growth history.

Despite the large number of studies on individual GCs in the Galaxy, and on the Galactic GC system as a whole, we still lack a definite picture of their formation, and of their initial properties, such as  the initial mass function of the GC system, and its initial metallicity distribution \citep{lamers17, forbes18, renaud18}. This lack of knowledge of the properties of the original GC system makes difficult to reconstruct its subsequent evolution, since the parameter space of possible initial conditions  to explore -- initial masses, concentrations, spatial distribution,.. --  is large, and its subsequent evolution may depend on these unknown conditions. The picture is also complicated by the fact that  in the Milky Way~(hereafter MW), as well as in other galaxies, there is evidence that not all GCs formed {\it in-situ}, that is in the galaxy itself. Part of the current population may indeed have been formed in satellite galaxies, subsequently accreted \citep{bellazzini03, forbes10, leaman13, mackey10}. Satellite accretions might enlarge the initial population of GCs in a galaxy  \citep{tonini13, 2017MNRAS.465.3622R}, but also  kinematically heat the pre-existing {\it in-situ} population  \citep{kruijssen15, 2016MNRAS.459.2905M}, similarly to the case of {\it in-situ} disc stars. Those stars  -- repeatedly heated by accretion events - increase their random motions, contributing to the thick disc/-stellar halo field populations \citep[see, for example, ][]{qu11, jeanbaptiste17}.

Once considered as single stellar populations, made of stars with the same initial chemical composition and same age, GCs have revealed in the last decade to be much more complex than what previously thought. Old and massive clusters show the presence of multiple stellar populations, with large star-to-star variations in  light element abundances, that have been detected both photometrically and spectroscopically, and whose origin is currently debated \citep{gratton12, bastian17}. Even more striking is the evidence that some Galactic GCs show large metallicity spreads, not compatible with the possibility that these clusters formed in a single burst of star formation. Some of the GCs with large metallicity spreads -- like $\omega$ Centauri  \citep{johnson10} -- have been suggested to have an extragalactic origin. \citet{bekki03}, for example, proposed that  $\omega$~Cen may constitute the remnant~(or nucleus) of a disrupted dwarf galaxy accreted by the MW. Another possibility is that Galactic GCs with metallicity spreads may have been formed by mergers between parent GCs once located within dwarf spheroidal galaxies, then accreted by the MW \citep{vandenbergh96,bekki03, 2016MNRAS.461.1276G}.
The hypothesis of GCs mergers taking place in dwarf galaxies -- and not in the MW itself -- is usually preferred since the probability of mergers and collisions among the current population of Galactic  GCs is generally very low, taken into account the current volume  they cover, and their velocity dispersions \citep{vandenbergh96, 2016MNRAS.461.1276G}. However, if this extra-galactic scenario may be valid for some GCs \citep{vandenbergh96}, it is hardly reconcilable with the properties of another Galactic GC: Terzan~5. This bulge/inner disc cluster indeed has been shown to contain a multimodal metallicity distribution function~\citep{massari14}, covering the whole extent of metallicity of the Galactic bulge, and also a significant age spread  \citep{origlia11, ferraro16}. Its main populations lie on the same age-metallicity relation described by bulge stars in the field  \citep{ferraro16}, thus making an extragalactic origin for this cluster very unlikely.

In this paper we aim at starting to explore the evolution of a system of disc GCs, evolving in a MW-like potential. The reasons to study the population of disc GCs are numerous. First, we know that the current population of MW GCs is not only distributed into the Galactic halo: the population of metal-rich clusters~(i.e. $\approx \rm{[Fe/H]} > -1$~dex) is indeed more concentrated and flatter than that of the more metal-poor GCs - and is indeed associated to the thick disc and bulge populations \citep{armandroff88, armandroff89, zinn85, zinn90, zinn91, minniti95, barbuy98, cote99, ortolani99, 1999AJ....117.1792D,1999AJ....117..277D,2007AJ....134..195C, vandenbergh03, bica06, bica16}. Second, even if nowadays this population constitutes only about one third of the total GC population \citep[][2010 edition]{harris96}, there are reasons to think that the population of disc GCs was more numerous in the past. Former GCs member stars have been indeed recently discovered in the inner regions the Galaxy \citep{schiavon17, fernandez17}  and simulations also show that tidal effects in the inner kpc of a MW-like galaxy could have been efficient over time in reducing the initial population, destroying part of it \citep{2017MNRAS.465.3622R}. Third, part of the initial population of Galactic disc GCs have experienced kinematic heating over time due to accretion events, and be now found in the inner halo \citep{kruijssen15}, while at high redshifts it may have been formed in a disc. Finally, GCs with kinematics and spatial properties of disc populations exist also in nearby galaxies, as in M31 \citep[see, for example][]{morrison04, caldwell16}.

For all these reasons, in this paper we want initiate a new study of the dynamical evolution of a system of disc GCs. A number of works have investigated this topic, by modelling the evolution of single globular clusters in MW-like potentials, exploring a range of initial masses, concentrations, and orbital parameters for the clusters \citep{berentzen12, rossi15a, rossi15b, rossi16, rossi17}. Other works have studied the evolution of a whole population of open clusters in the disc of a spiral galaxy~\citep{2012MNRAS.427L..16F,2017MNRAS.464.3580D}. The novelty of our work is to present an $N$-body model of the self-consistent evolution of a whole population of disc GCs, thus naturally taking into account their gravitational mutual interactions. For short-term evolution of thin disc star clusters such approach have been used in several studies  \citep[see e.g.,][]{2012MNRAS.427L..16F,2017MNRAS.464.3580D}. More precisely, in this work, in this work, we present an $N$-body simulation of a GC system containing $128$ resolved clusters with initial masses of $10^7$~\Msun, and whose evolution is followed for $1.5$~Gyr. We will show that under certain conditions --  namely an initially numerous and/or massive enough GC disc population -- mergers, fly-bys  and mass exchanges between GCs can occur over time, at a rate of few events for Gyr. In a forthcoming study~(\paptwo), we will complement this analysis with a detailed study of the long-term evolution of the remnants of GC mergers, by means of direct $N$-body simulations.  The paper is organized as follows: in Section.~\ref{sec::model}, we introduce the Milky Way-like potential adopted in this paper,  the modelled GCs properties, and the code adopted to run the simulation. In Section~\ref{sec:analysis} we make use of  a simple analytic treatment to estimate the probability of physical collisions in a system of GCs, as a function of their number and spatial distribution. In Section~\ref{sec:results}, we present the results of the $N$-body simulation, focusing in particular on the mergers, fly-bys and related mass exchange occurring in our simulated GC system. Discussion and implications of our results are given in Section.~\ref{sec::conclusions}.  

\begin{figure*}
\begin{center}
\includegraphics[clip=true, trim = 7mm 2mm 17mm 5mm, width=0.33\linewidth]{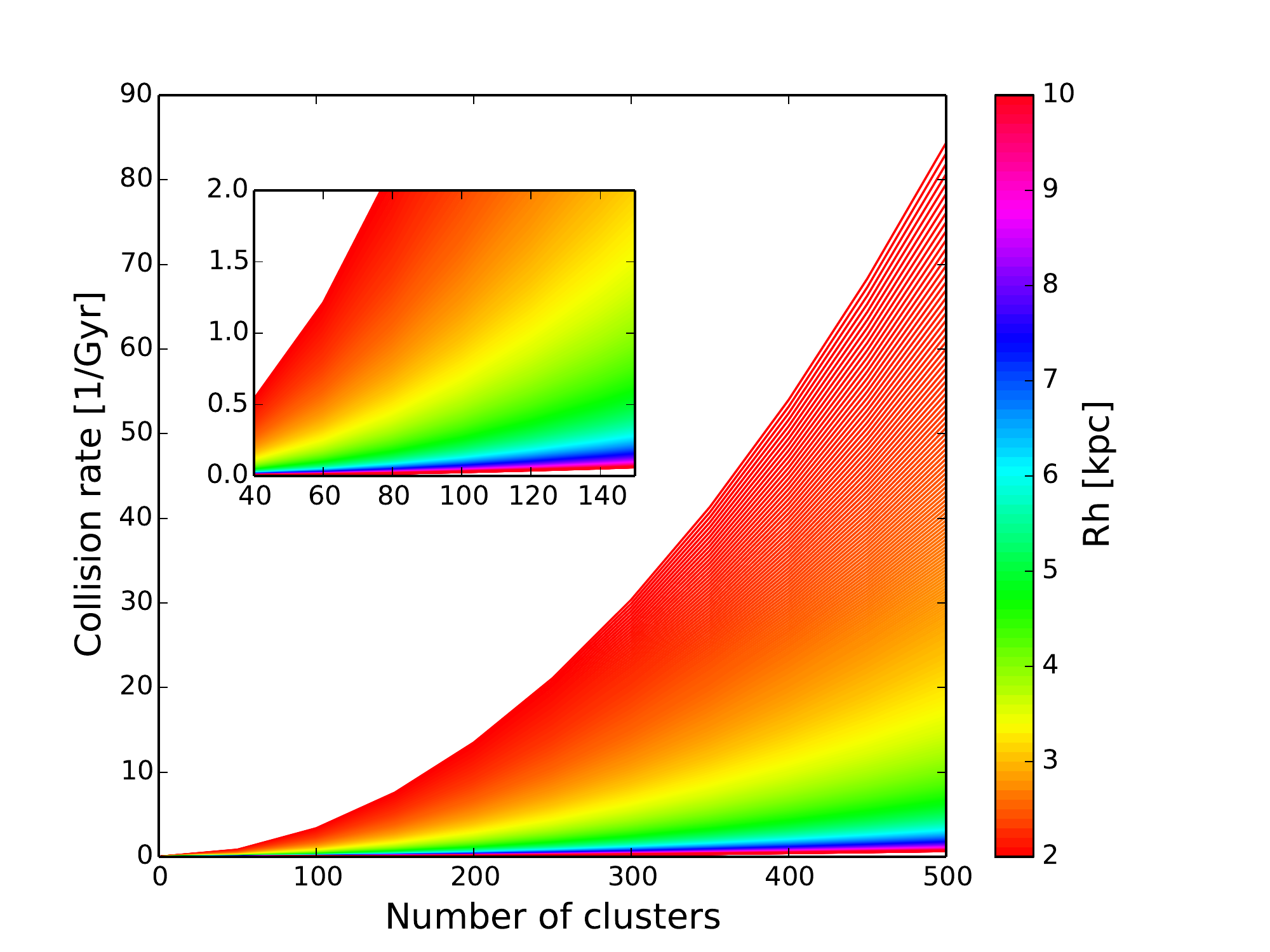}
\includegraphics[clip=true, trim = 7mm 2mm 17mm 5mm, width=0.33\linewidth]{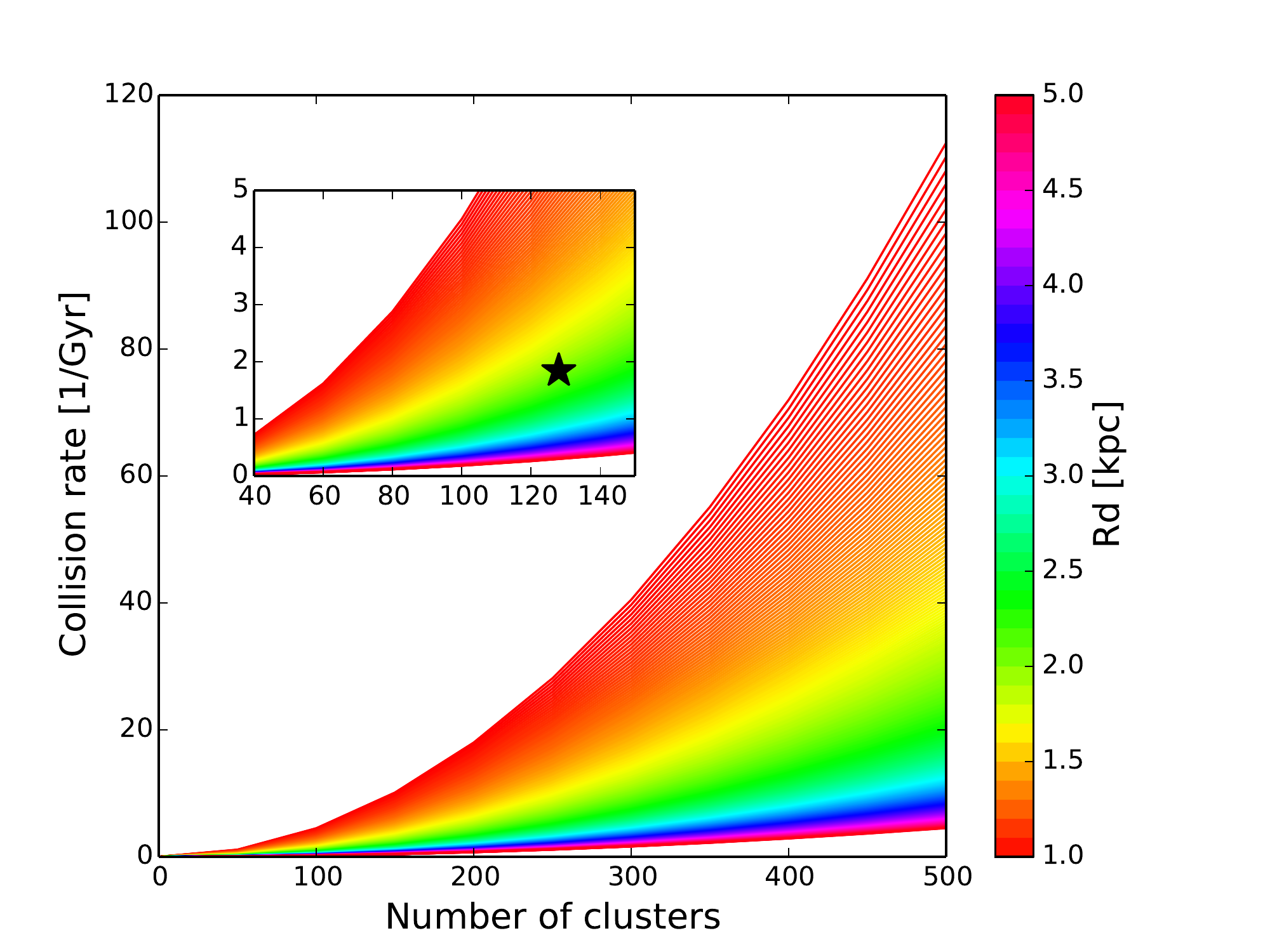}
\includegraphics[clip=true, trim = 7mm 2mm 17mm 5mm, width=0.33\linewidth]{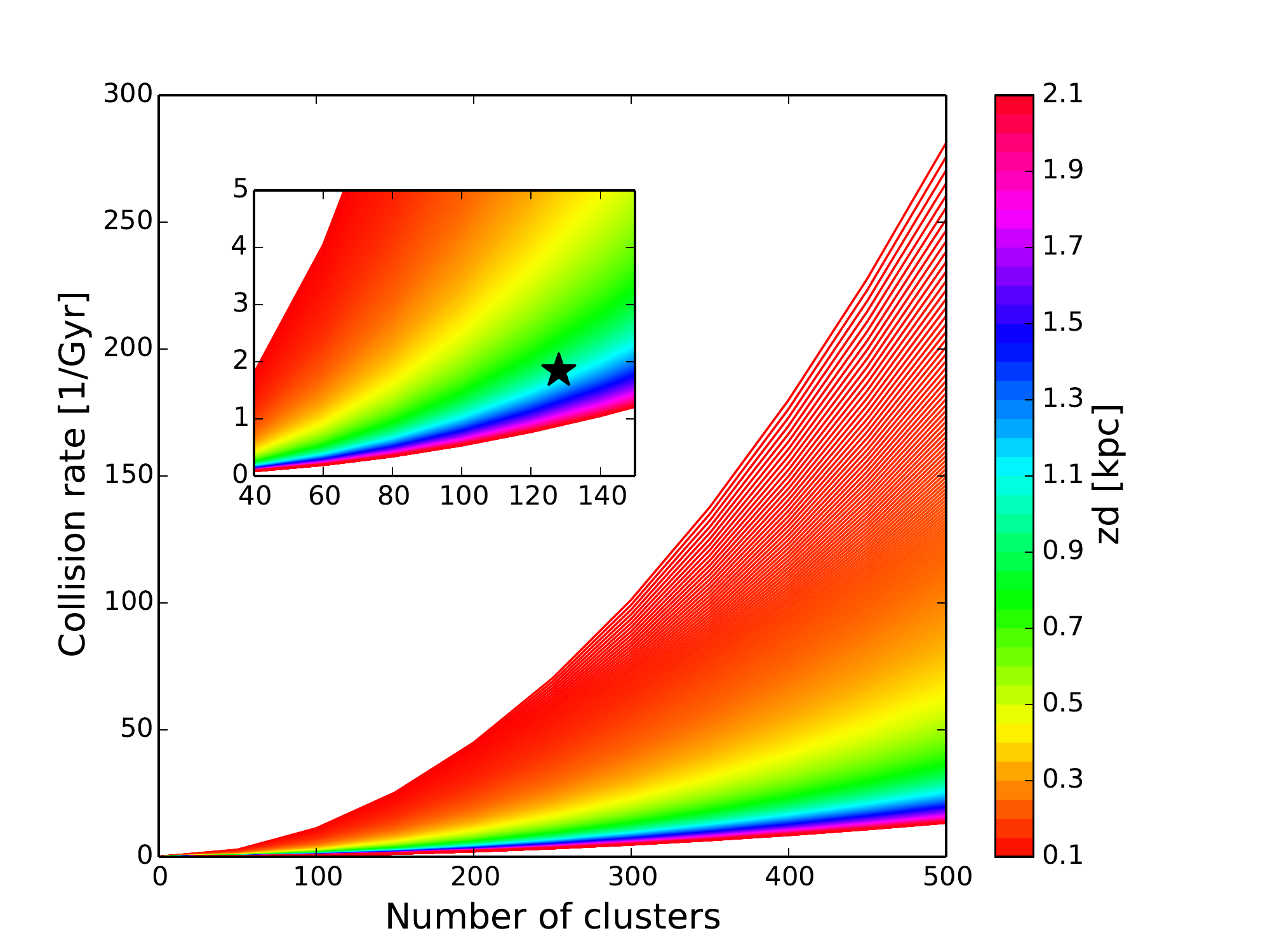}
\end{center}
\caption{Cumulative collision rate as a function of the number of globular clusters, in the case of a homogeneous spherical distribution~(left panel), and of a disc distribution~(middle and right panels). The color codes the size of the spherical halo~(left panel), the radial scale length of the disc~(middle panel) for a fixed scale height of $0.8$~kpc, and the vertical scale height~(right panel) for a fixed scale length of $2$~kpc.  In each plot, the inset shows a zoom for a number of clusters between $40$ and $150$. In the inset in the middle and right panels, the asterisk indicates the collision rate corresponding to a disc of  scale length of $2$~kpc and scale height of 0.8~kpc.}\label{fig::collisions}
\end{figure*}

\begin{figure*}
\begin{center}
\includemedia[width=1\hsize,activate=onclick,
 addresource=mov_swf/evolution_gcc00y_short.mp4.swf,flashvars={source=mov_swf/evolution_gcc00y_short.mp4.swf
 &loop=false &scaleMode=letterbox &autoPlay=false &controlBarMode=floating &controlBarAutoHide=false}]{\includegraphics[width=1\hsize]{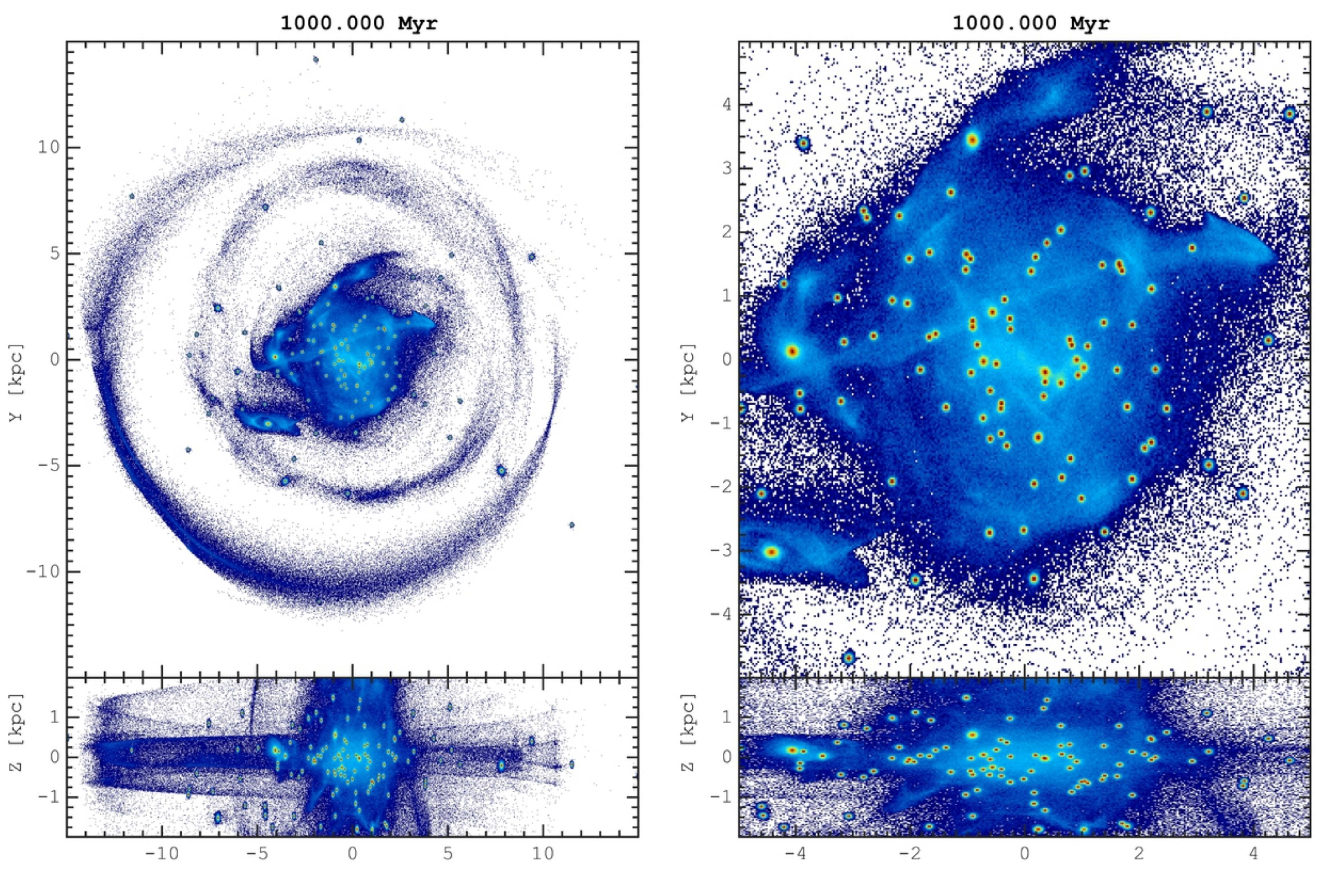}}{mov_swf/evolution_gcc00y_short.mp4.swf}
\mediabutton[mediacommand=test:playPause]{\fbox{Play/Pause}}
\caption{Stellar density distribution~(in logarithmic scale) of the entire globular cluster system in our simulation.  Projections on the $\rm xy$ and $\rm xz$ planes are shown. In the right panels, a zoom in the inner disc regions is shown. Bound globular cluster stars are clearly seen as red over-densities, while light/dark blue regions correspond to stars escaped from the clusters and lost in the field. The high-resolution animation of the evolution of the simulated GC system can be found here: \mywebpage.}
\end{center}\label{fig::global_evolution}
\end{figure*}

\section{Model}\label{sec::model}
To follow the internal physics of individual GCs, and the evolution of a GC system as a whole, we choose to  simulate the dynamics of resolved, $N$-body, GCs in an analytic and time-independent  MW potential. The advantage of this approach is to allow the study of the mutual interactions and gravitational effects among simulated GCs, together with their internal evolution, mass gain and loss, all these effects ``embedded`` in the overall gravitational field exerted by a MW-type galaxy. The limits of this approach are the lack of consideration of the time dependent variations and inhomogeneities of the galactic gravitational potential. Both these effects are very likely concurring to shape the evolution of the GC system over time \citep[see, for example, ][]{kruijssen15, 2017MNRAS.465.3622R}.
 
More specifically, in this work we simulate the dynamics of $128$~disc GCs, each cluster being represented by $10^5$ point masses, accounting for a total of $1.28\times10^7$ particles. We chose to use a number of disc GCs, $N_{GC}$, greater than the current one in the MW -- approximately equal to 50\footnote{This may still be a lower limit to the number of disc/bulge GCs in the MW. Recently, indeed, $22$~new GC candidates have been discovered in the inner MW \citep{minniti17}, with approximately half of them having $\rm [Fe/H] > -1$~dex.} -- under the hypothesis that this latter value current number of Galactic disc GCs is not representative of the initial one. This choice is justified by the fact that tidal effects, early disruption, and heating in the halo may have all contributed to reduce the primodial disc population. The exact choice of the initial number  $N_{GC}=128$ is arbitrary and dictated by the fact that this number allows a particularly effective algorithm for MPI-parallelization. 

Each GC is initially represented by a King model with core radius $r_c = 4$~pc and tidal radius of $r_t = 80$~pc, corresponding to a dimensionless central potential $W_0 = 6$ \citep{binney87}. Each cluster is made of  particles all having the same mass $m_*=100~\rm M_{sun}$. Their initial positions and velocities, in the absence of the external gravitational galactic potential, have been generated with the NEMO software~\citep{1995ASPC...77..398T}.

The initial positions and velocities of the system of disc GCs have been calculated by using the method described in~\citet{2009MNRAS.392..904R}. This consists of an iterative method to construct equilibrium phase models of stellar systems, and it can be adapted to a number of different mass distributions with arbitrary geometries, satisfying different kinematical constraints. The population of $128$~GCs is at equilibrium in the galactic gravitational potential, that is represented by an axisymmetric composite disc, made of a  thin and a thick component, and a spherical dark matter halo. The spatial distribution of this GCs population follows a Miyamoto-Nagai density profile~\citep{1975PASJ...27..533M} with a characteristic radial length of 2 kpc and a vertical height of 0.8 kpc, similar to those of the Galactic thick disc in Model II of~\cite{pouliasis17}. As shown by these authors, the adopted parameters satisfy a number of observational constraints of the current MW: stellar densities at the solar vicinity, thin and thick disc scale lengths and heights, rotation curve, and the absolute value of the perpendicular force as a function of distance to the galactic centre. No specific kinematical constraint has been introduced to generate the initial velocities of this GCs population.  Note that the GCs~(especially those in the innermost kpcs) are not on circular orbits. The total mass of the GCs system is equal to $1.28\times10^9$~\Msun, but - as we will discuss later - this mass diminishes with time, as a result of the evolution  of the GCs with the overall Galactic tidal field, and of  GC-GC collisions.

To model the evolution of the GC system in this potential, we employ a $N$-body code~\citep{2014JPhCS.510a2011K}, in which gravitational forces are calculated using an MPI-based parallel hierarchical tree method including Intel's Advanced Vector Instructions~(AVX). We use an opening angle parameter $\theta = 0.5$. A Plummer potential is used to soften gravitational forces, with a constant softening length of $\epsilon = 0.2$~pc. The equations of motion are integrated using a leapfrog algorithm with a fixed time step of $10^4$~yr. The Tree Code works efficiently with the system we simulate, because of the sparse spatial distribution of GCs in the MW-like potential, while the internal evolution of GCs is performed with a direct summation accuracy. We performed several test simulations of a single isolated GC, finding that the choice of our Tree Code parameters and numerical resolution guarantees that the single GC is at equilibrium over timescales longer than those simulated in this work~($\approx1.5$~Gyr), with a typical relative error in energy conservation of $1-2\%$ at the end of the simulation.

\section{On the probability of physical collisions in a system of disc globular clusters}\label{sec:analysis}
Before moving to show and discuss the results of our $N$-body simulations, we present here some general considerations on the probability of physical collisions and mergers in a GC system. These considerations constitute some of the main motivations to the work presented in this paper, and provide the reasons to investigate numerically the collisions in a system of disc GCs.

The majority of the current population of Galactic GCs is associated to the  stellar halo, both because of  their individual metallicity~([Fe/H]$<-1$~dex for about 2/3 of the Galactic GC population, see \citet{harris96}) and current positions \citep{harris96} as well as orbital parameters, when available \citep{dinescu97, dinescu99, dinescu99b, dinescu03, dinescu07, dinescu10, dinescu13, chemel18}.

Given a number $N_{GC}$ of GCs, redistributed over a volume $V$, and having all identical characteristic radii $R_{GC}$, the mean free path of a cluster in the system is given by:
\begin{equation}
\lambda=\frac{1}{n\Sigma}
\end{equation}
with $n$ being the cluster volume density, that is $n=N_{GC}/V$ and $\Sigma$ the effective collision area, that is $\Sigma=\pi(2R_{GC})^2$.
The associated rate of physical collisions of a cluster with another one is thus $r=v/\lambda$, where $v$ is the typical velocity of the cluster in the system, while $R$ is the cumulative rate of collisions over the $whole$ cluster population, scales quadratically with $N_{GC}$, being 
\begin{equation}\label{eq::rate}
R = N_{GC} r = N_{GC}^2 v \Sigma/V.
\end{equation}

For instance, for a population of a hundred halo GCs, distributed inside a sphere of radius $7$~kpc, similar to the half-mass radius of the current population of Galactic halo~(i.e. metal-poor, [Fe/H]$<-1$~dex) GCs, with a typical velocity of $100$~km/s, the cumulative collision rate -- would be less than $0.1$~Gyr$^{-1}$. In this estimate, we assume that all GCs have a physical size of $3$~pc, a typical value for the core radii of the current GCs in the Galaxy.
Of course, as shown in Fig.~\ref{fig::collisions}~(left panel) and in Eq.~\ref{eq::rate}, this rate is both a function of the number of clusters in the system, and of the spatial extent of the population. As an example, to have at least one collision per  Gyr in a population of a hundred GCs redistributed in  a spherical halo, the spatial extent of this population should be of about $3$~kpc, thus significantly smaller than the current extent of the system of halo GCs in the Galaxy.
However, for the same number of GCs, the rate of collisions is significantly higher if this population is redistributed in a disc, rather than in a halo. As an example, for a system of  $N_{GC}=128$ GCs, redistributed in a disc whose scale-length and scale-height are respectively equal to $2$~kpc and $1$~kpc -- thus similar to the scale-length and height of the Galactic thick disc \citep{bovy12, bovy16} -- the rate of physical collisions would be $1.8$~Gyr$^{-1}$. As well as for the halo, also for a population of disc GCs, this rate is both a function of the number of GCs, and of their radial and vertical extent, as shown in the middle and right panels of Fig.~\ref{fig::collisions}. In all cases, if this population is redistributed in a compact, flattened configuration -- as it is the case of the current population of metal-rich GCs in the Galaxy, \citep[see, for example][]{armandroff89} -- and if this population is numerous enough, $1-2$ collisions per Gyr can be expected.
In other words, collisions and mergers of GCs could have taken place also {\it in-situ}, i.e. in our Galaxy, and not necessarily only in dwarf galaxies then accreted into the Milky Way~\citep{vandenbergh96,2016MNRAS.461.1276G} provided that, at early epochs, the Galactic GC system was numerous and compact enough. In the following, we indeed show by means of our $N$-body simulation, that such a collision rate is realistic, and that indeed  gravitational interactions can take place in a system of disc GCs having similar characteristics in terms of number of clusters, scale length and height, leading to mergers, fly-bys and mass exchanges. 

\begin{figure}
\centering
\includegraphics[width=1.\hsize]{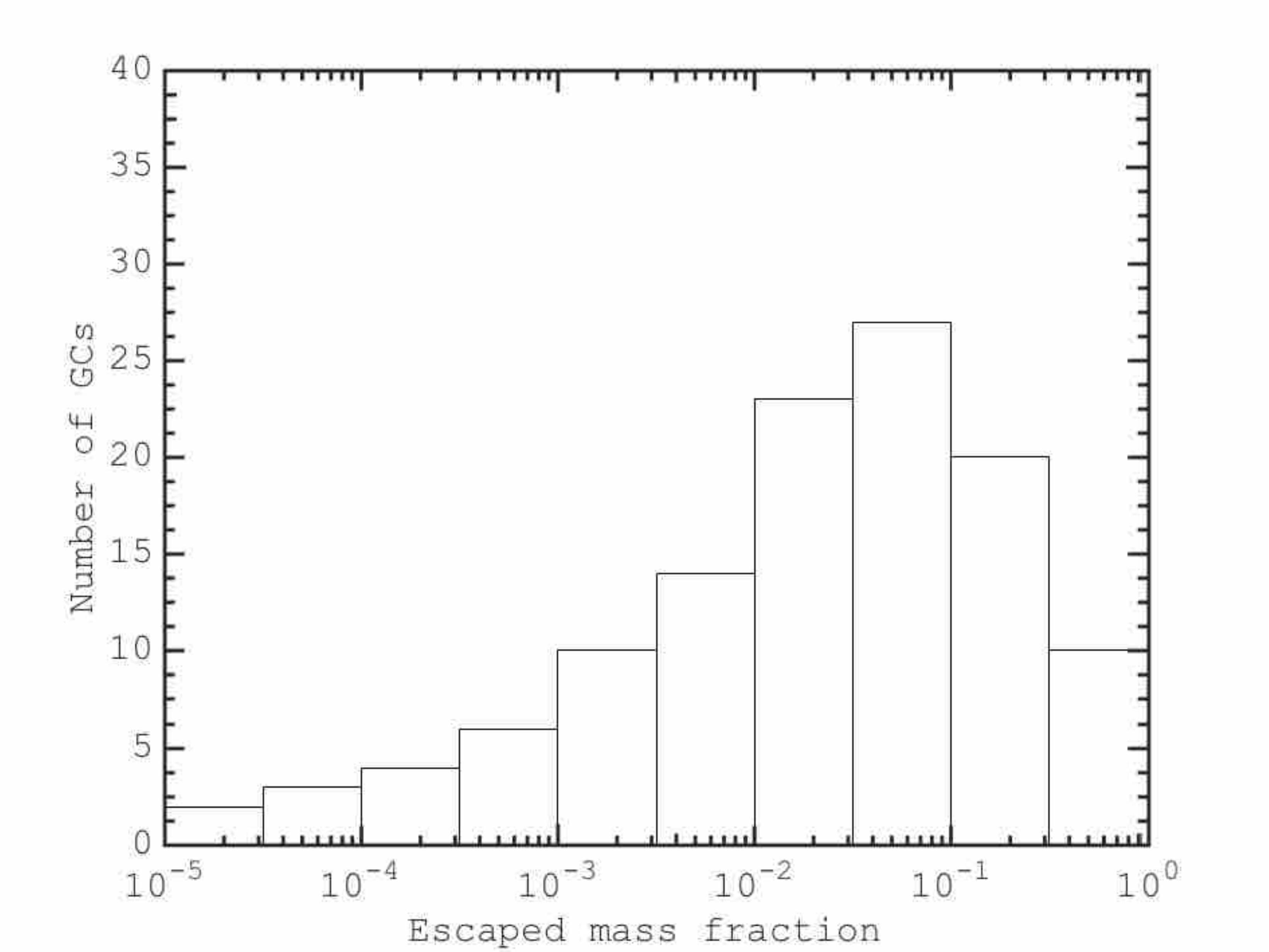}
\includegraphics[width=1.0\hsize]{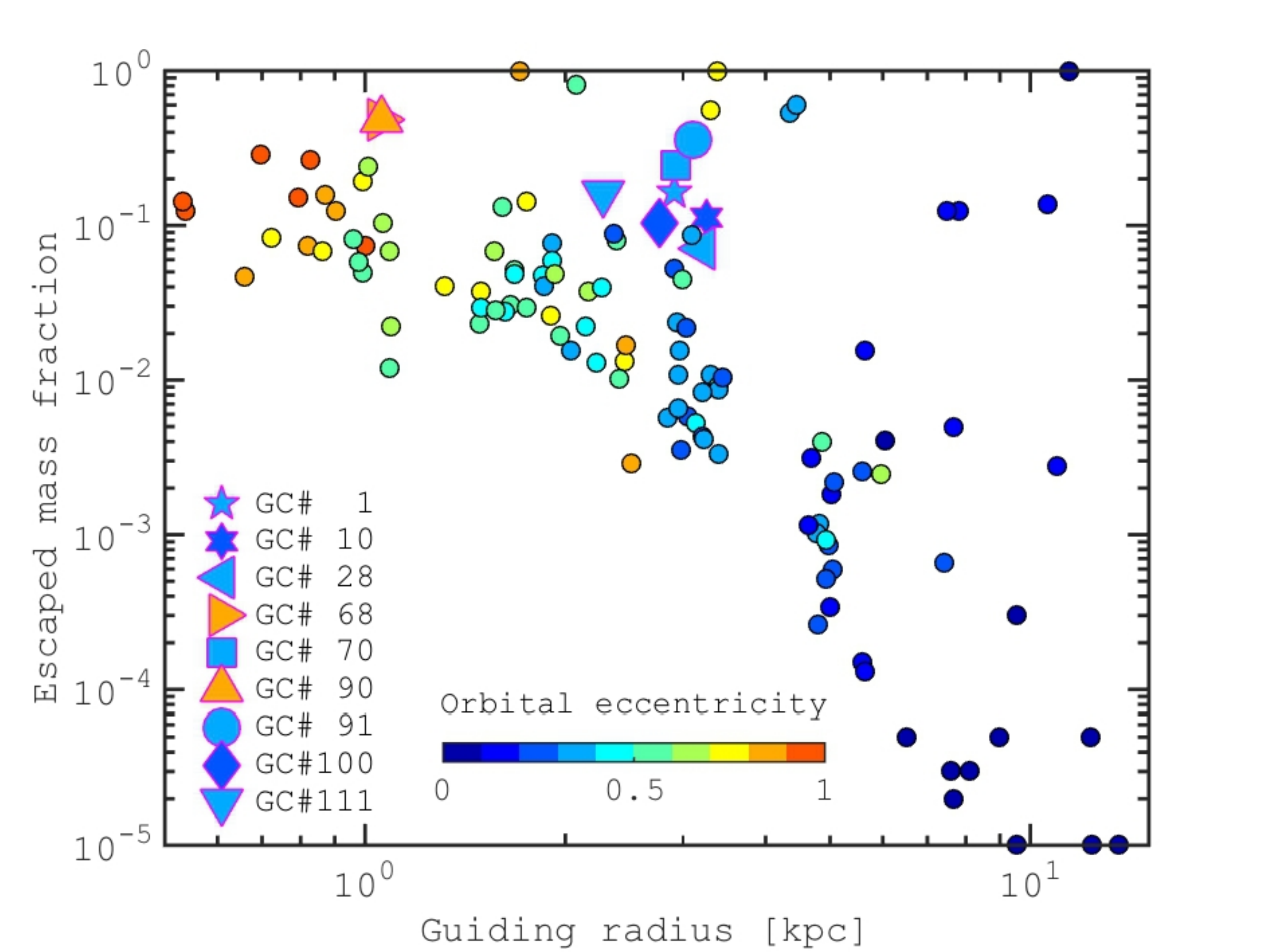}
\caption{ {\it Top panel:}  Distribution of the escaped mass fraction from the simulated GCs, after $1.5$~Gyr of evolution. {\it Bottom panel:} Escaped mass fraction as a function of the orbital guiding radius. Colour code depicts the GCs orbital eccentricity.}\label{fig::stat}
\end{figure}

\section{Evolution of a $N$-body system of disc GCs: Results}\label{sec:results}

\subsection{Unbound stars, major mergers and mass accretions: some definitions}\label{sec:global_evolution}
The stellar density distribution of the GC system, after $1$~Gyr of evolution in the galactic potential described in Sect.~\ref{sec::model}, is shown in Fig.~2. Most of the GCs are redistributed in the inner few kpc of the Galaxy, with only few clusters  at distances greater than $r=5$~kpc from the Galactic centre. 
In this figure, GCs appear as density peaks of the stellar distribution, embedded in a low density field of stars escaped from the clusters over time. This loss of stellar mass from the clusters is the consequence of the action of the galactic gravitational field and of the mutual gravitational interactions among GCs, as we will discuss in more details in \paptwo. The effect of the galactic tidal field is also responsible of generating extended tidal tails around clusters, which are probes -- on large scales -- of the GC orbits~\citep{montuori07}. Note that the evaporated stars are distributed in a thick disc-like structure and  contribute significantly to the bulge region, where populations of disrupted GCs have been recently observed in the Milky Way inner disc~\citep[see e.g., ][]{2017MNRAS.466.1010S, 2017ApJ...846L...2F, 2018ApJ...862L...8L}. To quantify the number of mergers and the amount of mass exchange between clusters, we define the following quantities.
{\it The escaped mass} is defined as the difference between the gravitationally bound  mass  of the GC at any given time $t$, and its initial mass. \\
{\it Bound stars} are defined as all stars inside the cluster tidal radius which have a negative energy:
\begin{equation}
E = E_{\rm kin} + E_{\rm pot} < 0\, \label{eq::EE}
\end{equation} 
where $E_{\rm kin}$ and $E_{\rm pot}$ are the kinetic and potential energy of a star. 

Finally, following \cite{2009MNRAS.392..969J},  the {\it cluster tidal radius} $r_t$ is defined as 
\begin{equation}
\Oo r_t = \left( \frac{G M}{(4-\beta^2 )\Omega^2} \right)^{1/3}\,,
\end{equation}
where $M$ is the mass of the cluster, $\beta = \kappa/\Omega$ is the normalized epicyclic frequency , $\Omega$ is the angular speed of the cluster on a circular orbit at its guiding radius, and $V(R)$ is the value of rotation curve at a distance $R$ from the centre. In our simulations typical value of the tidal radius varies in the range of $100-300$~pc, while the core density is in the range of $(5-10)\times 10^3~\Msun~pc^{-3}$.

Over the first $1-1.5$~Gyr of evolution, most of the GCs have maintained the majority of their initial mass, and only $14$~clusters, corresponding to 10\% of the GCs in the system, have lost more than 20\% of their initial mass, as shown in Fig.~\ref{fig::stat}~(top panel). For the same orbital guiding radius, GCs that experience a strong mass loss tend to have slightly higher orbital eccentricities~(shown by colour), as shown in Fig.~\ref{fig::stat}~(bottom panel). Globally, of the initial mass $M_{GC}=1.28\times10^{9}~\Msun$ in the GC system, $9.6\%$ is unbound and dispersed into the field after $1.5$~Gyr.  Fig.~\ref{fig::stat}, bottom panel, shows that the amount of mass loss from a GC depends on its guiding radius, the lower the GC guiding radius, the higher the fraction of stellar mass lost over time. In particular this plot shows the existence of a zone of avoidance, with the escaped mass fraction being always higher than a minimum value which depends on the GC guiding radius.

We have checked that this trend is not due to an initial internal relaxation of our GCs - since we do not see any rapid expansion of the simulated GCs once placed in the galactic field. It is rather a consequence of GC-GC collisions - whose rate is higher among clusters having small guiding radii --  and of the significant variations in the tidal field experienced by clusters with high orbital eccentricities, which are preferentially found in the inner few kpc of the galaxy. Both these effects contribute to determine the trends observed in Fig.~\ref{fig::stat}.

Together with the effect of the mean galactic gravitational field, the GC system evolves also through the effect of mutual GC interactions. We assume that a GC contains an external population if stars from another, or several, cluster(s) are located within its tidal radius and if their energies satisfy the condition shown in Equation~\ref{eq::EE}. We refer to the GC that has accreted mass from another GC as the host, while the accreted stars are referred to as the external population. 
We define a {\it major merger} between GCs as an accretion where two interacting GCs completely merge and, as a consequence, the external population has a mass comparable to the bound mass of the host. Accretions where the host contains at least $1\%$ of mass accreted from another GC, without any subsequent major merger,  are referred to as {\it mass exchanges or mass captures}.
In the following, we describe all individual cases of major mergers and mass capture in our simulated population of disc GCs. To identify mergers and interactions in the system,  we number our GCs from $1$ to $128$. Note also that the same GC can be host in an accretion event and concurrently part of its mass can contribute to the external population of another host GC. Therefore, a given GC can receive mass from another cluster, and at the same time, contribute with a fraction of its mass to a companion GC.

Note also a unique case of evolution of globular cluster \#38 in the outer disk. This cluster is located in the top right of Fig.~\ref{fig::stat} with the escaped mass fraction of unity. The cluster is fully disrupted at the galactic outskirts, and its remnant is recognized as a nearly circular stream at $R\approx12$~kpc at later times~(see Fig.~\ref{fig::global_evolution}). During the evolution of the system, this cluster interacted twice with different GCs. At $t=140$~Myr it was perturbed by GC\#24, and then GC\#38 started to expand and lose mass. Later at $t\approx0.5$~Gyr GC\#38 has flown through the potential well of the tail of another GC~(\#50) which speeded up the evaporation of stars and triggered the formation of the extended tidal tail.

\subsection{Major mergers and mass accretions}\label{sec::selection}

At the end of the simulation, several GCs in the system have experienced major mergers or mass accretion from other clusters, leading to GC remnants with composite populations. In particular, we find evidence of\footnote{In the following, we identify with~(H) the host GC.}:
\begin{itemize}
\item an early major merger, which leads to a GC remnant with equal fractional contributions from two parent clusters~(GCs~\#68 and \#90);
\item a major merger with a minor accretion from a third cluster~(GCs \#70(H), \#91 and \#100);
\item a host with two accreted populations~(GCs \#111(H), \#28 and \#91);
\item six clusters with accreted populations~(GCs \#001(H) and \#010; GCs \#010(H) and \#001; GCs \#024(H) and \#123; GCs \#028(H) and \#111; GCs \#080(H) and \#121; GCs \#100(H) and \#91, GCs \#123(H) and  \#024).
\end{itemize}

\begin{figure*}
\includegraphics[width=1\hsize]{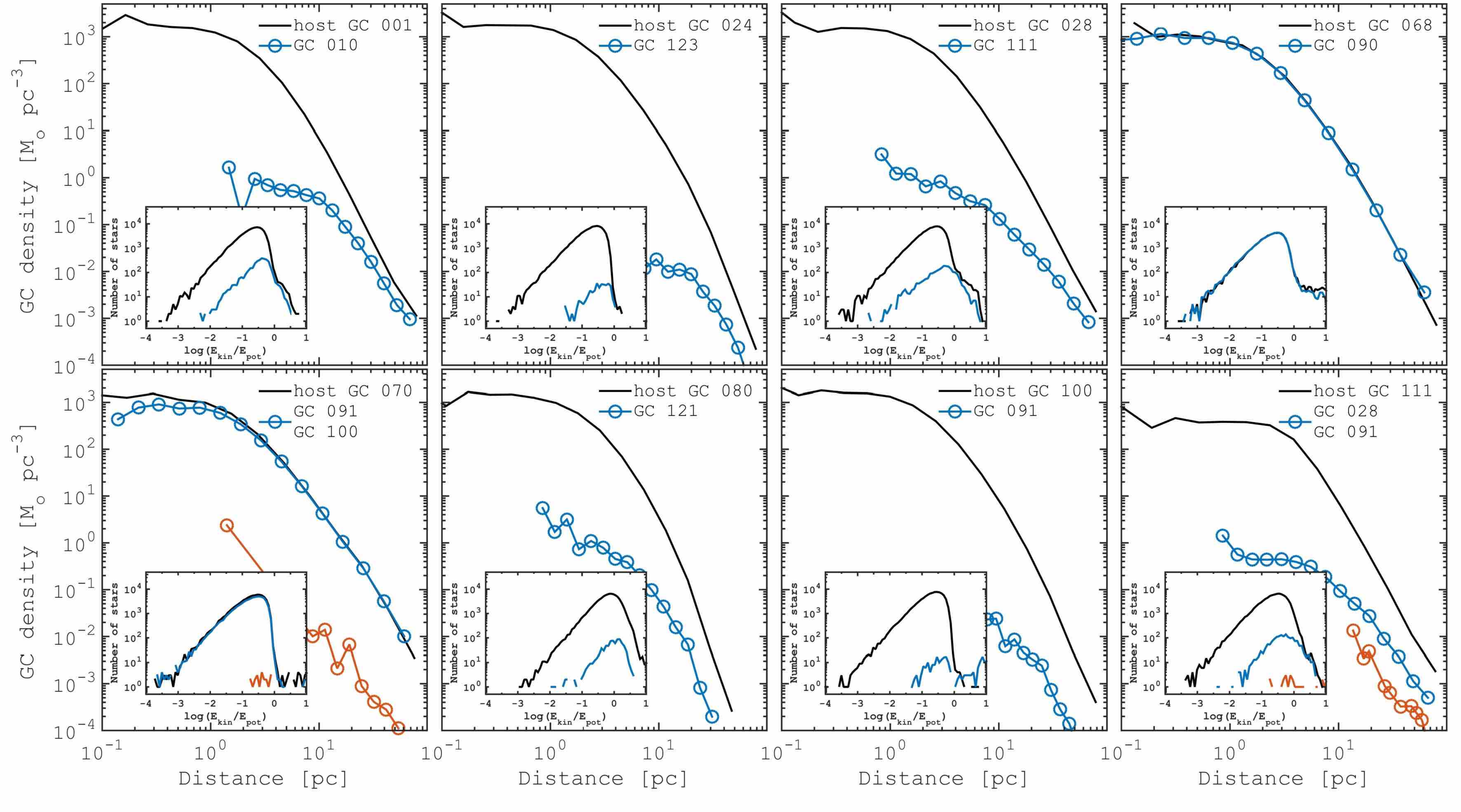}
\caption{Radial density profile in units of $\Msunpcc$ for each GC~(host) which have gravitationally bound~(see Eq.~\ref{eq::EE}) external stars inside of the tidal radius. In each subframe the distribution of the $\rm \log(E_{\rm kin}/E_{\rm pot})$ is shown. Solid black lines show the host cluster characteristics; external populations are shown by blue and red~(in case of two external populations) symbols.}\label{fig::density_energy}
\end{figure*}

In Fig.~\ref{fig::density_energy} we show the final volume density  of all simulated GCs, which have experienced one or several mergers/accretions. For each of these GCs, we also show the kinetic-to-potential energy ratio distribution for gravitationally bound populations within the tidal radius\footnote{ We have also checked that the results of the analysis we are to present are robust by measuring the parameters of accreted and in-situ populations within a sphere of 50~pc which is much smaller than a typical tidal radius.}: in-situ and accreted population(s). In the case of major mergers, one can see that the final density profile and the final energy distribution are equal for both populations, which are therefore fully mixed after $1-1.5$~Gyr of evolution. In the case of mass accretion from another GC, the accreted stars do not penetrate in the innermost regions of the host cluster -- because their kinetic energy is too large. Hence for the accreted populations, distance from the GC centre increases with their decreasing mass. For the long-term evolution of these accreted populations in the host GC we refer to \paptwo.
In the following, we describe in more detail the mergers and mass exchanges taking place in our simulated GC system.

\subsubsection{A major merger~(GCs 68-90)}\label{sec:major_merger}
One of the two major mergers involves a pair of clusters~(GCs~$\#68-\#90$) with an initial separation of about $\approx 150$~pc. These GCs are located in the inner disc, at a distance from the galactic centre smaller than $2$~kpc, and have small relative velocities in comparison to orbital motion. 
 
Fig.5~(top-left panel) shows the evolution with time of the spatial separation between the two GCs which merge after about $50$~Myr from the beginning of the simulation. The final  face-on and edge-on distributions of stars initially belonging to the two GCs are shown in the right panel of Fig.~5. The merger remnant -- corresponding to the density peak in the plot -- is surrounded by an extended distribution, or `halo` of unbound stars lost in the field. During the merger, both clusters have  lost around $20-25\%$ of their initial stellar mass, due to their mutual tidal effects, while the influence of  the galactic tidal field  causes further mass loss and, at the end of the simulation, the GC remnant has lost about half of its initial mass~(see Fig.~5, bottom-left panel). Nevertheless, the radial density profile of the GC merger remnant is that of a typical globular cluster~(see Fig.~\ref{fig::density_energy}, for GC~\#68 as a host). This new cluster will however consist of two fully mixed populations.

\begin{figure*}
\begin{center}
\includegraphics[width=0.49\hsize]{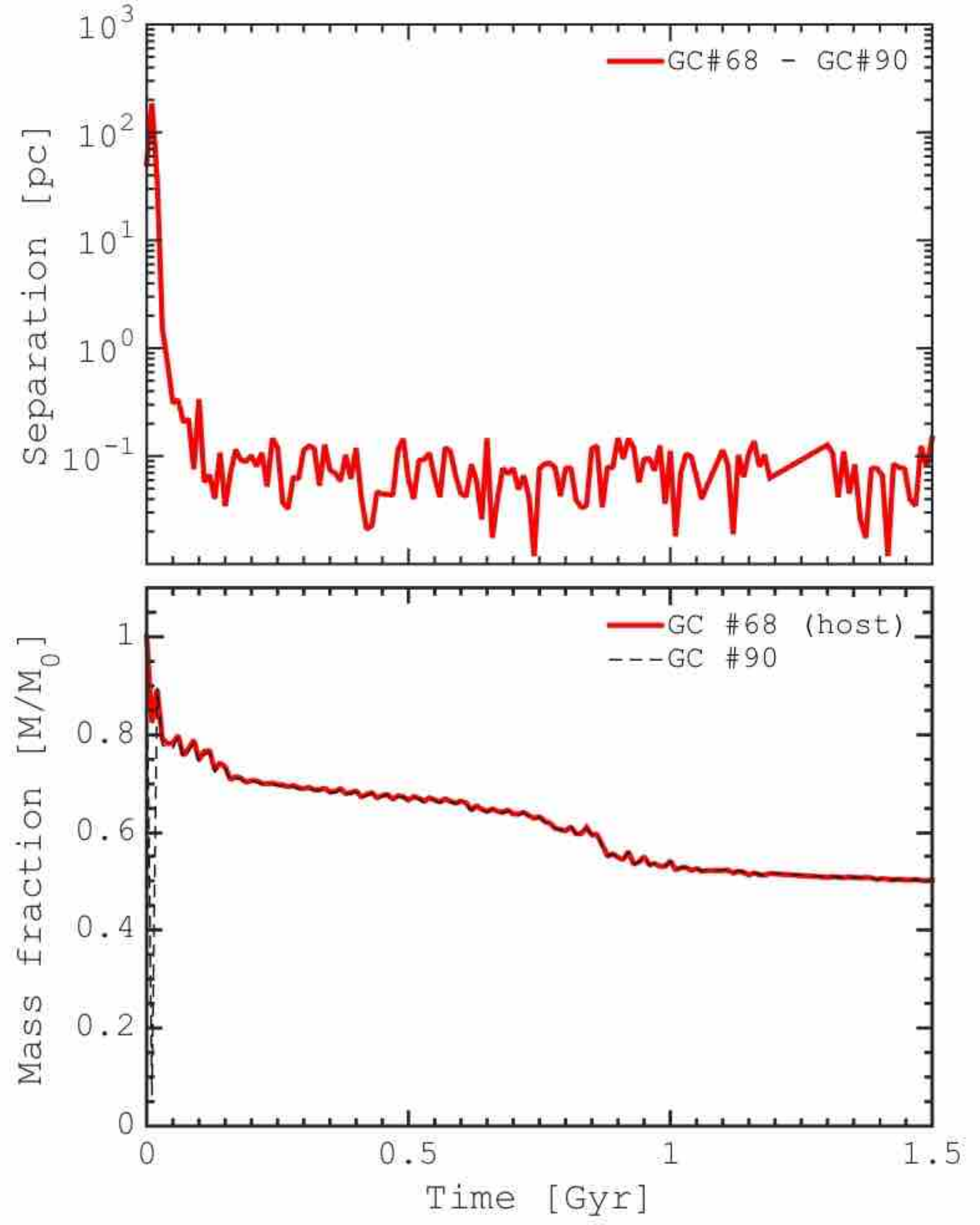}
\includemedia[width=0.49\hsize,activate=onclick, addresource=mov_swf/dynamics_68_90.mp4.swf, flashvars={source=mov_swf/dynamics_68_90.mp4.swf
 &loop=true &scaleMode=letterbox &autoPlay=false &controlBarMode=floating &controlBarAutoHide=false}]{\includegraphics{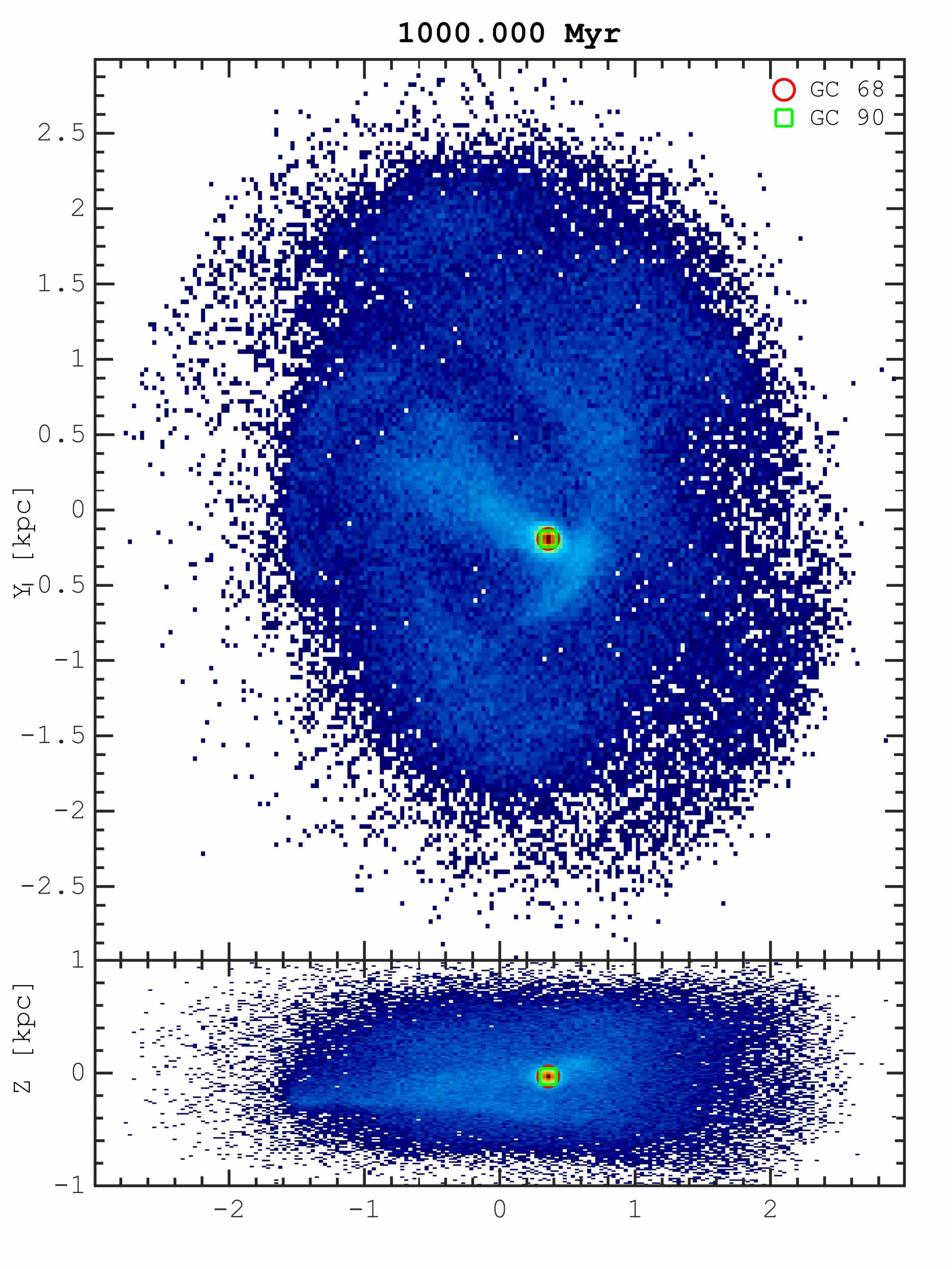}}{mov_swf/dynamics_68_90.mp4.swf}
 \mediabutton[mediacommand=test:playPause]{\fbox{Play/Pause}}\caption{Left: evolution of the separation between GCs 68 and 90~(top), and evolution of gravitationally bound mass~($E_{\rm kin}/E_{\rm pot}<1$) enclosed inside the $50$~pc from the host~(GC 68) globular cluster center. Right: the same as for the global evolution plot~(Fig.~2), but only two GCs, 68 and 90, are shown. The high-resolution animation can be found here: \mywebpage.}
\end{center}\label{fig::sm68-90}
\end{figure*}

\subsubsection{The case of a mass accretion, followed by a major merger~(GCs 70-91-100)}\label{sec:minor_major_merger}
\begin{figure*}
\begin{center}
\includegraphics[width=0.49\hsize]{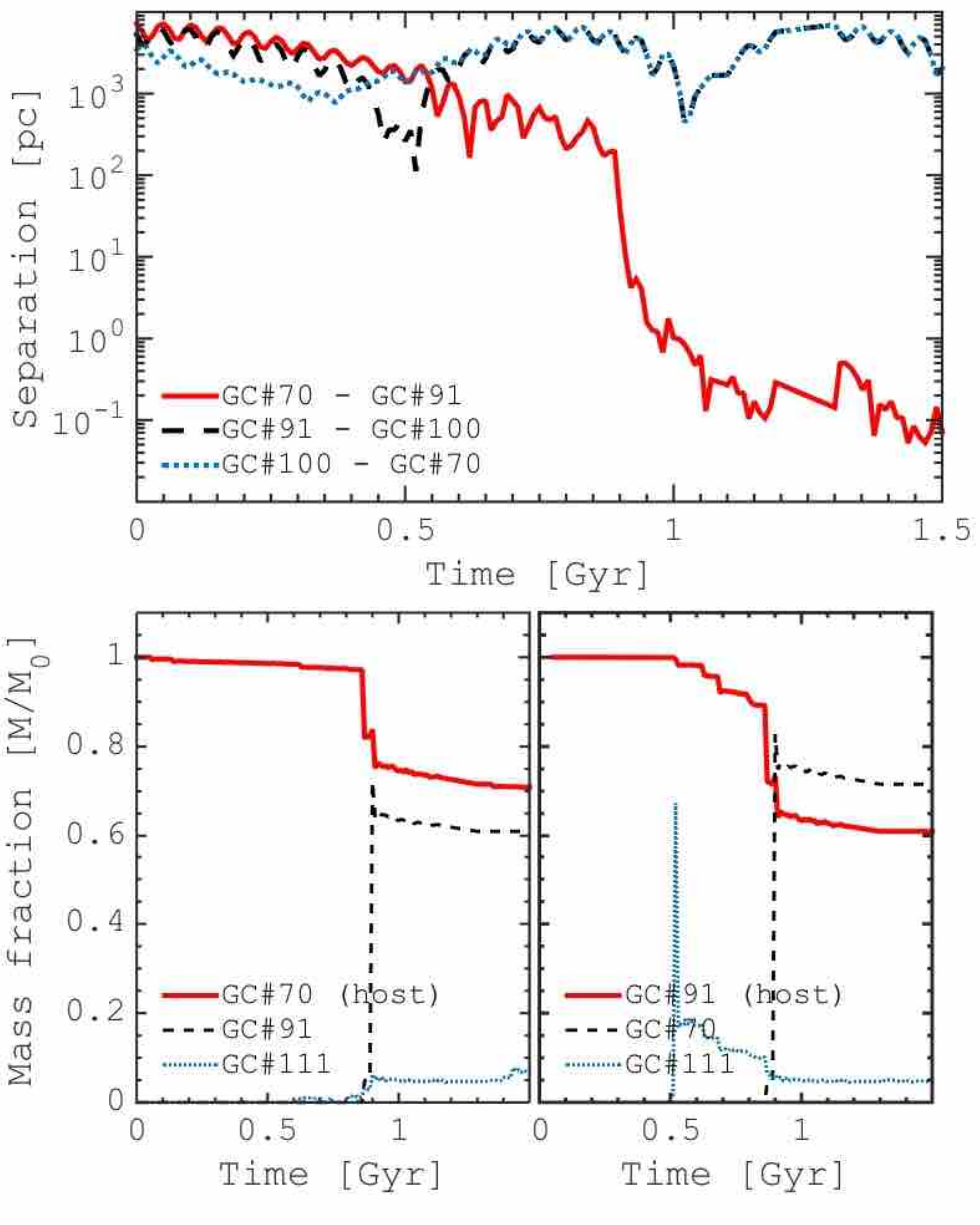}
\includemedia[width=0.49\hsize,activate=onclick, addresource=mov_swf/dynamics_70_91_100.mp4.swf, flashvars={source=mov_swf/dynamics_70_91_100.mp4.swf
 &loop=true &scaleMode=letterbox &autoPlay=false &controlBarMode=floating &controlBarAutoHide=false}]{\includegraphics{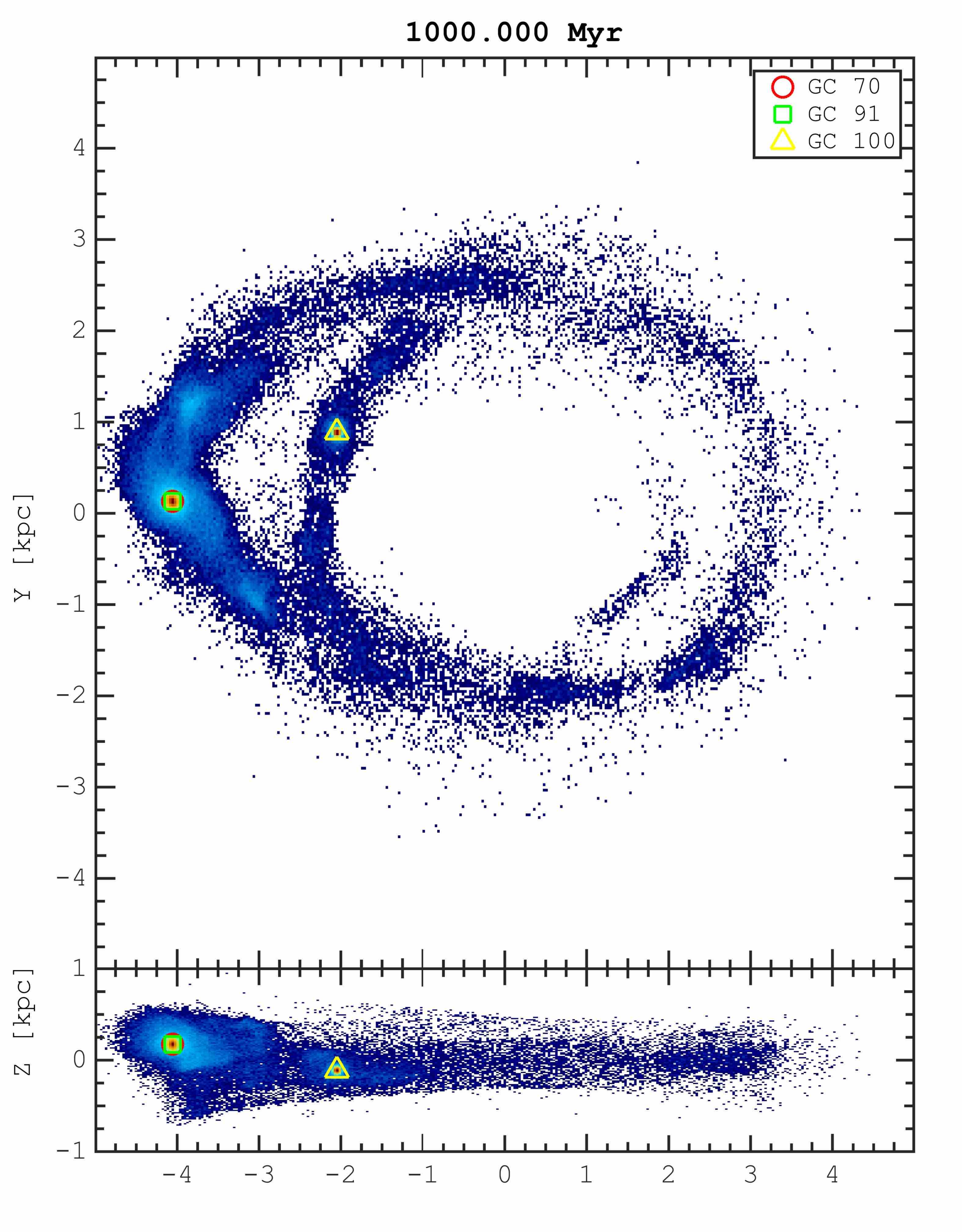}}{mov_swf/dynamics_70_91_100.mp4.swf}
 \mediabutton[mediacommand=test:playPause]{\fbox{Play/Pause}}
\caption{As in Fig.~5, but for GCs 70, 91 and 100.}\label{fig::sm91-70-100}
\end{center}
 \end{figure*}

Another interesting case we obtain in our simulation is the interaction between multiple GCs, resulting in a remnant GC consisting of  three gravitationally bound populations. In the top panel of Fig.~\ref{fig::sm91-70-100}, we show the evolution of the separation between the three pairs of clusters. The contribution of each progenitor to the mass fraction of the two GCs involved in the major merger~(GCs~$\#70-\#91$), is illustrated in the bottom panel of Fig.~\ref{fig::sm91-70-100} as a function of time. The three clusters initially move on similar orbits -- in terms of spatial extension and eccentricity -- and cross their paths several times during the simulated time interval. 
At $t \approx 0.5$~Gyr, during a close passage, the GC $\#91$ captures $15\%$ of the mass from the GC $\#100$. Then at a later time~($\approx 0.8$~Gyr) this same cluster~($\#91$) fully merges with another one~(\#70). During this merger, a significant fraction of the external population accreted at $t=0.5$~Gyr from GC~$\#100$ is lost in the field, and the resulting system contains $\approx 70\%$ of mass from the GC~$\#70$, $\approx 60\%$ from the GC~$\#91$ and only a small percentage, $\approx 1-2$\%, from the cluster GC \#100. Repeated interactions like this one are possible because of the similar orbital motions of the clusters that imply small relative velocities. The final density profiles of the two clusters involved in the major merger~(GCs \#70 and \#91) are similar, while the minor population~(GC\#100) shows a power-law distribution and is found only outside the core of the GC remnant, at $r>1$~pc~(see Fig.~\ref{fig::sm91-70-100}).

\subsubsection{An example of multiple tidal mass capture~(GCs 111-28-91)}

\begin{figure*}
\begin{center}
\includegraphics[width=0.49\hsize]{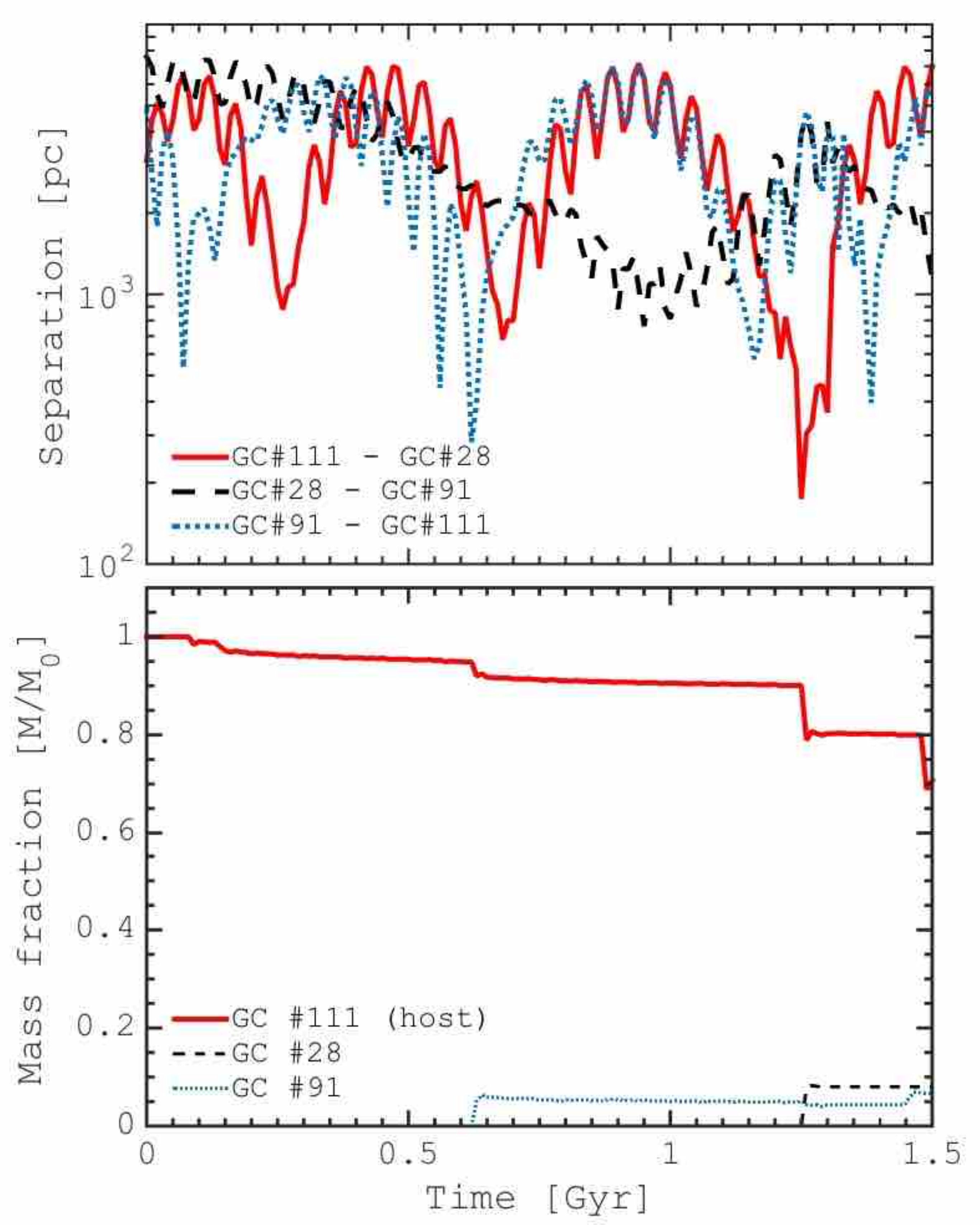}\includemedia[width=0.49\hsize,activate=onclick, addresource=mov_swf/dynamics_111_28_91.mp4.swf, flashvars={source=mov_swf/dynamics_111_28_91.mp4.swf
 &loop=true &scaleMode=letterbox &autoPlay=false &controlBarMode=floating &controlBarAutoHide=false}]{\includegraphics{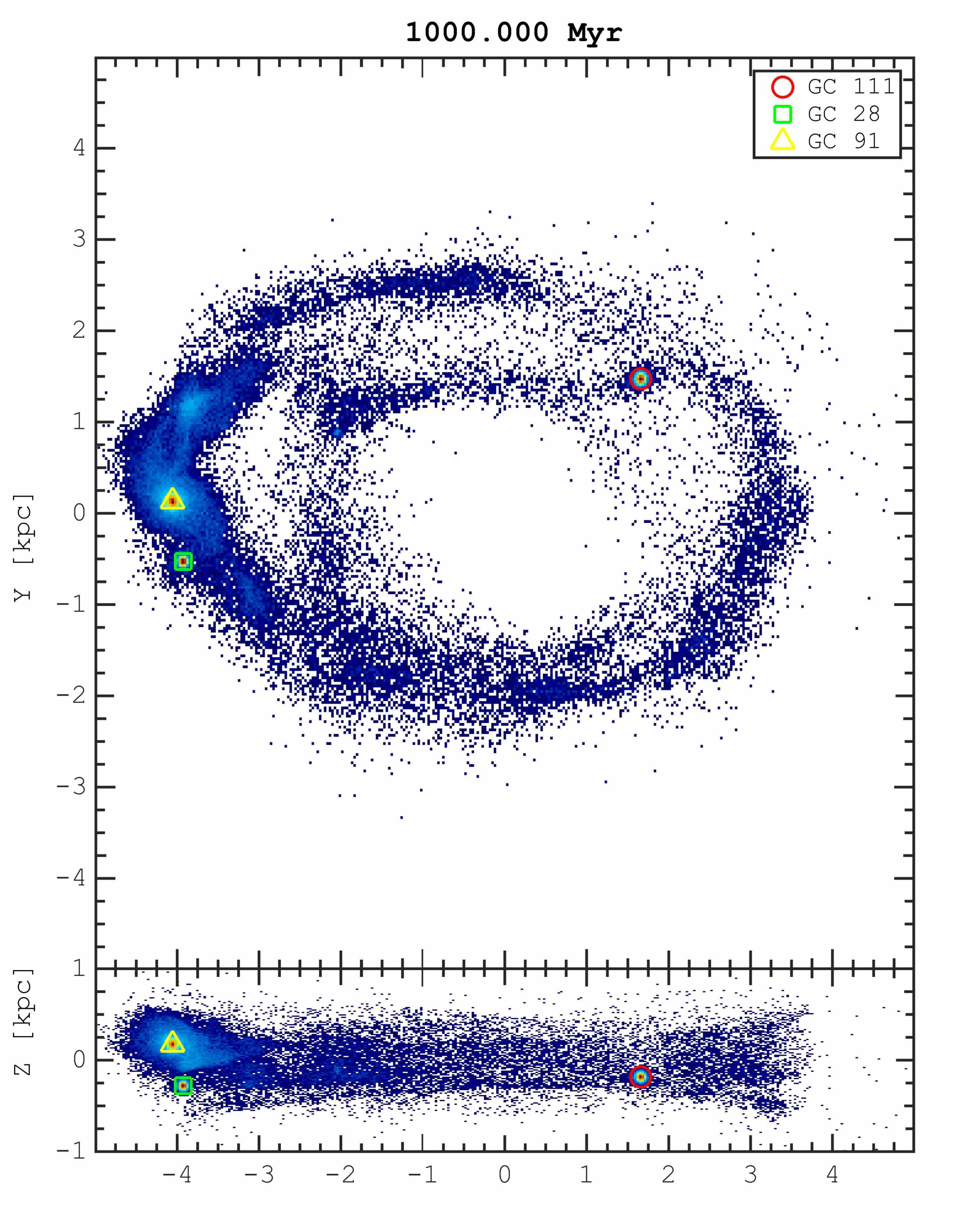}}{mov_swf/dynamics_111_28_91.mp4.swf}
 \mediabutton[mediacommand=test:playPause]{\fbox{Play/Pause}}
\caption{As in Fig.~5, but for GCs 111, 28 and 91.}\label{fig::sm111-28-91}
\end{center}
\end{figure*}

Beside the two major mergers described above, at the end of the simulation we find also several cases of GCs containing few percent of mass accreted from other clusters in the system, during their orbital evolution in the galaxy. All cases are discussed in  Appendix~\ref{app}, while here we are interested in describing the case of the interaction between the GCs \#111, \#28 and \#91. In Fig.~\ref{fig::sm111-28-91} we show the evolution with time of the relative separation of these clusters with respect to the host GC\#111. This cluster initially accretes $1-2$\% of mass from the GC~\#91, after about $t=0.6$~Gyr from the beginning of the simulation. Then, at the final stages of the simulation, during a close passage another $1-2$\% of mass is accreted from the from GC $\#28$. Thus, even if at the final time the remnant cluster is for the vast majority made of stars originated in the GCs\#111, in its outer regions~(see Fig.~\ref{fig::density_energy}) we find a contamination, at a few percent level, from clusters which have passed close to it. Relative separations of the order of few hundred pc between these GCs are sufficient to generate these mass exchanges. Note that the mass exchange is also mutual, because when the GC~\#111 accretes mass from the GC~\#28, it also provides this with latter few percent of its original mass.

\section{Discussion and Conclusions}\label{sec::conclusions}
In this work we have studied the dynamical evolution of a system of disc globular clusters in a Milky Way-like, axisymmetric potential, by means of  a $N-$body simulation run for $1.5$~Gyr.  The main results of our work and the predictions of our model can be summarized as follows:
\begin{itemize}
\item By means of analytic calculations of physical collisions rate, we show that 1-2 collisions per Gyr are expected for a system of globular clusters redistributed in a galactic disc, with scale length/height similar to those of the current MW thick disc.  Thus collisions and mergers of GCs could have also taken place in-situ, i.e. in our Galaxy, and not necessarily only in dwarf galaxies then accreted into the Milky Way. 
\item In $N$-body simulations we have shown that, for a disc population of GCs with scale length and height compatible to those of the Galactic thick disc, mergers among GCs can take place at a rate of $\sim 1$ per Gyr. For a similar  merger rate, the GC population must be numerous~($\sim100$GCs) and the clusters need to be massive~(the initial mass adopted for all clusters in our simulation being equal to $10^7 M_\odot$). 
\item While orbiting in the disc, the mutual gravitational interaction can lead to mass exchanges and captures among simulated GCs. We indeed find several cases where a GC accretes few per cent of its final mass from one or several companions. 
\item In the case of a major merger between two GCs, the final density profiles are equal for both GC populations, indicating that they can be fully mixed after $\sim  1$~Gyr  after the merger.  
\item In the case of mass capture during GCs fly-bys,  the accreted mass tends to redistribute in the outer regions of the host GC, because the kinetic energy of the accreted stars is usually too high to allow them to penetrate in the innermost GC regions.
\end{itemize}

Note that these findings extend the earlier works by Capuzzo-Dolcetta and collaborators~\citep{2006ApJ...644..940M, 2008ApJ...681.1136C, 2008MNRAS.388L..69C, 2012ASPC..453..237M, 2012ApJ...750..111A, 2013MmSAI..84..167C, 2014MNRAS.444.3738A}, who showed that mergers of GCs can occur in the nuclear regions of galaxies, i.e. few tens of parsecs,  leading to the formation of nuclear star clusters. Here we show that, in a population of disc globular clusters, mergers can occur also at distances of few kpcs from the galactic center,  and that the remnant of these mergers - or mass exchanges during fly-bys - are still  stellar systems that can be classified as GCs, on the basis of their morphology and kinematics, as we will show in a follow-up paper~(\paptwo). Before concluding we would like to briefly comment on the pertinence of the assumptions done  for our $N$-body model.  Because the conditions at early epochs of the GC system and of the Galaxy are still largely unknown, it is hard to say how realistic our assumptions can be. The adoption of a time-independent potential like that of \citet{pouliasis17} that we employed here -- which was developed as an axisymmetric mass model of the current Milky Way -- is a first description of the Galaxy, and does not take into account that its properties at high redshift may have been significantly different from the current ones~(both in terms of masses of the different disc components and dark matter, and of its equilibrium state). 
The current population of disc/bulge  GCs in the Galaxy, defined as the GCs with $\rm [Fe/H]> -1$~dex,  consists of approximately 50 members \citep[][2010 edition]{harris96}, however their census may still be largely incomplete, as shown for example by the recent discovery of a dozen of new metal-rich GCs toward the bulge \citep{minniti17}. Moreover, part of this initial population of disc GCs may have been destroyed over time by tidal effects and/or kinematically heated during the accretion of satellites and now be part of the Galactic halo GC population \citep{kruijssen15}. Thus, assuming a disc population which contains larger~(factor of $2-3$) than the current number of Galactic disc GCs may still be reasonable for the GC system at high redshift, but this value is currently still difficult to probe.
Finally, since the $N$-body model presented in this work is dissipationless, units can easily be re-scaled and GCs masses can be reduced, for example, by a factor of two -- thus allowing to employ initial GC masses possibly less extreme than the values adopted in this work. Of course this rescaling must be adopted also for the Galactic potential. A stellar mass of the Galactic disc which is half of its current value would be compatible with that estimated for MW disc at redshift $z\sim 1$ \citep{snaith14}.

To conclude, this work suggests a possible {\it in-situ} process for explaining the existence of GCs like Terzan~5, containing a significant metallicity spread and several peaks in its metallicity distribution function.  Assuming that the progenitors of Terzan~5-like GCs  have a different chemical composition and even different ages, the  resulting GC remnant will exhibit several peaks or spread in chemical abundances.
At least the metal-poor peaks can be the signature of mergers and mass capture of former disc/bulge GCs that have taken place over time in the innermost region of the Galaxy. 
Also, this work predicts that some GCs in the Galaxy should contain a small percentage of stars -- typically few percent -- with metallicities different  from that of the large majority of the GC stars, and that this contamination can be the result of mass capture from companion clusters. Looking for the existence and recurrence of these contaminants in the population of Galactic GCs may help providing some constraints on the number of close encounters experienced by GCs over time, and thus indirectly, on their initial number and/or masses. 

\begin{acknowledgements}
The authors thank Florent Renaud for a careful reading of the manuscript and an anonymous referee for useful comments. This work was granted access to the HPC resources of CINES under the allocation 2017-040507~(Pi : P. Di Matteo) made by GENCI.  This work has been supported by ANR~(Agence Nationale de la Recherche) through the MOD4Gaia project~(ANR-15-CE31-0007, P.I.: P. Di Matteo). Numerical simulations are partially carried out using the equipment of the shared research facilities of HPC computing resources at Lomonosov Moscow State University supported by the project RFMEFI62117X0011 and the RFBR grant~(16-32-60043). AMB acknowledges support by Sonderforschungsbereich~(SFB) 881 `The Milky Way System' of the German Research Foundation~(DFG). PDM thanks the Max Planck Institute f\"ur Astronomie for their hospitality during several visits while this Paper was in progress.
\end{acknowledgements}

\bibliographystyle{aa}
\bibliography{references}

\begin{appendix}\label{app} 
\section{GCs interactions: minor mass capture}

\begin{figure*}
\begin{center}
\includegraphics[width=0.49\hsize]{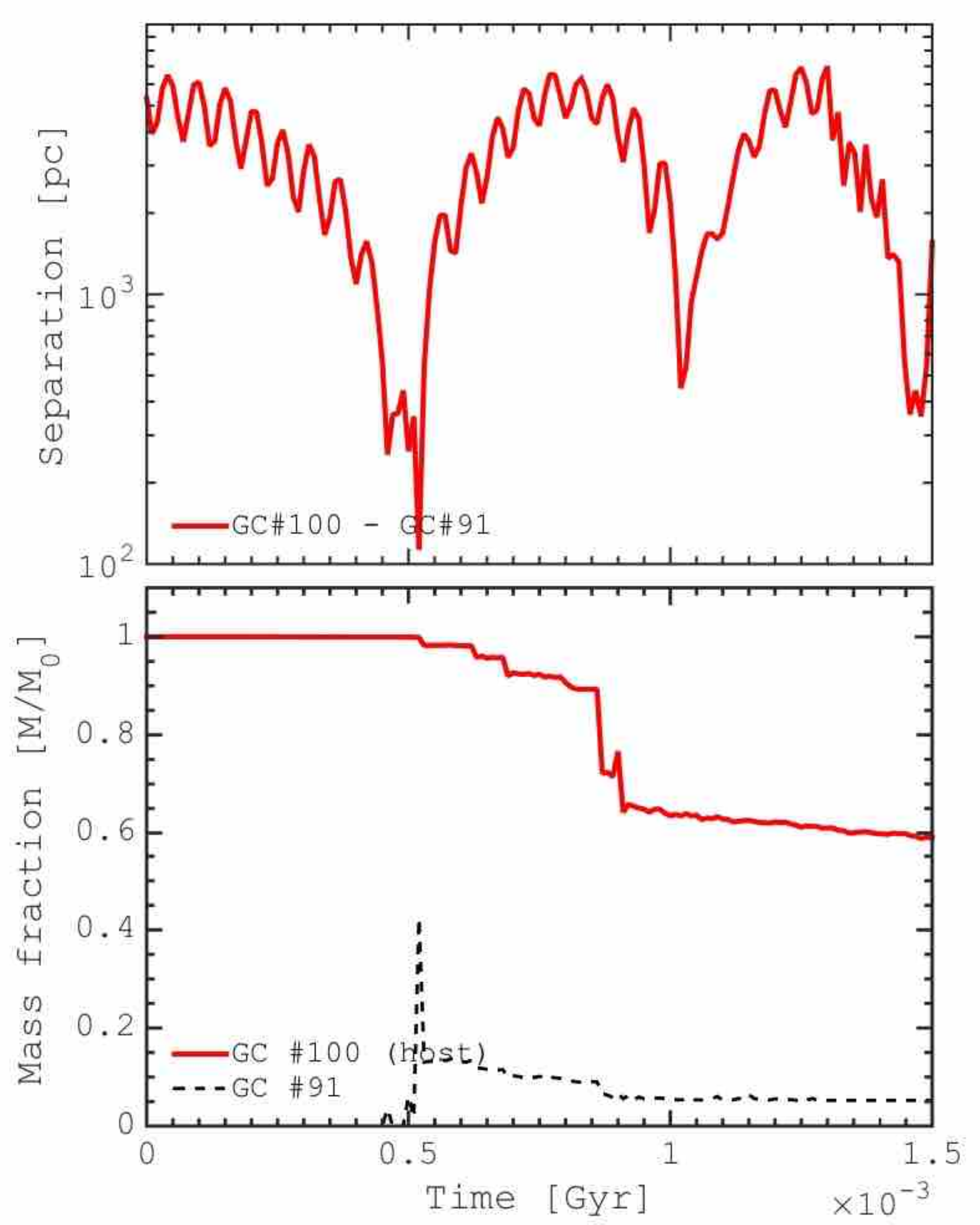}
\includemedia[width=0.49\hsize,activate=onclick, addresource=mov_swf/dynamics_91_100.mp4.swf, flashvars={source=mov_swf/dynamics_91_100.mp4.swf
 &loop=true &scaleMode=letterbox &autoPlay=false &controlBarMode=floating &controlBarAutoHide=false}]{\includegraphics{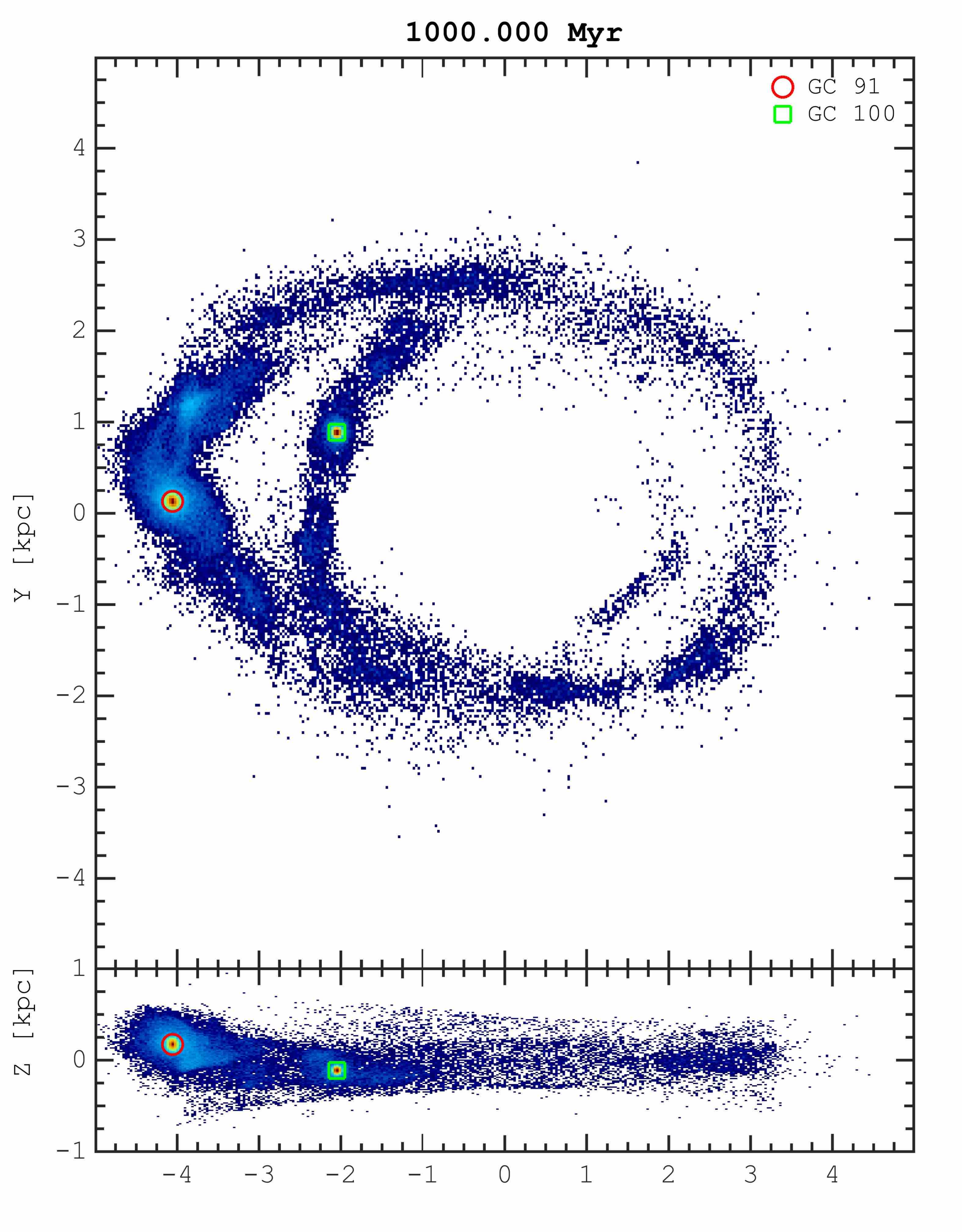}}{mov_swf/dynamics_91_100.mp4.swf}
 \mediabutton[mediacommand=test:playPause]{\fbox{Play/Pause}}\caption{As in Fig.~5, but for GCs 100 and 91.}\label{fig::sm91-100}
 \end{center}
\end{figure*}

\begin{figure*}
\begin{center}
\includegraphics[width=0.49\hsize]{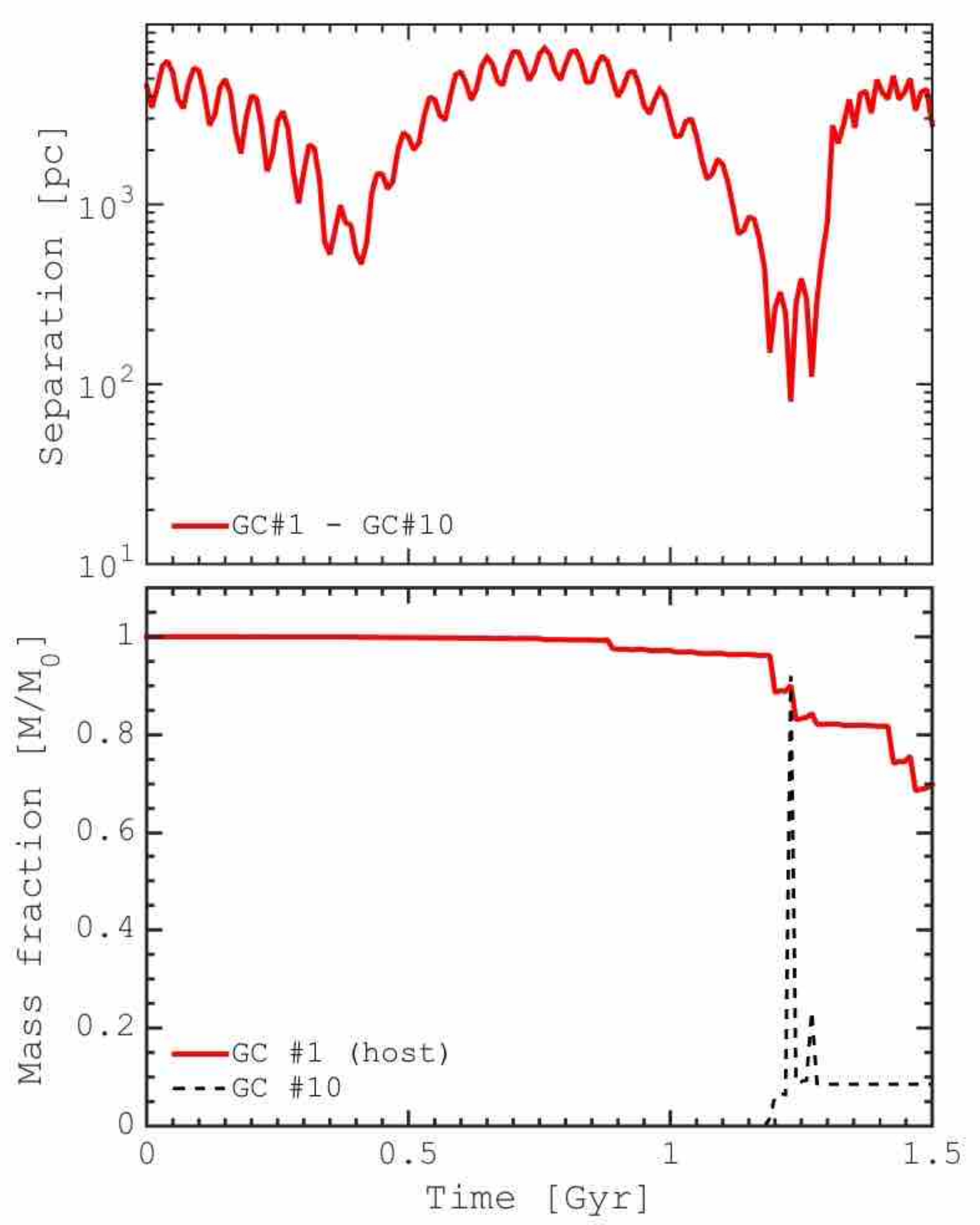}
\includemedia[width=0.49\hsize,activate=onclick, addresource=mov_swf/dynamics_1_10.mp4.swf, flashvars={source=mov_swf/dynamics_1_10.mp4.swf
 &loop=true &scaleMode=letterbox &autoPlay=false &controlBarMode=floating &controlBarAutoHide=false}]{\includegraphics{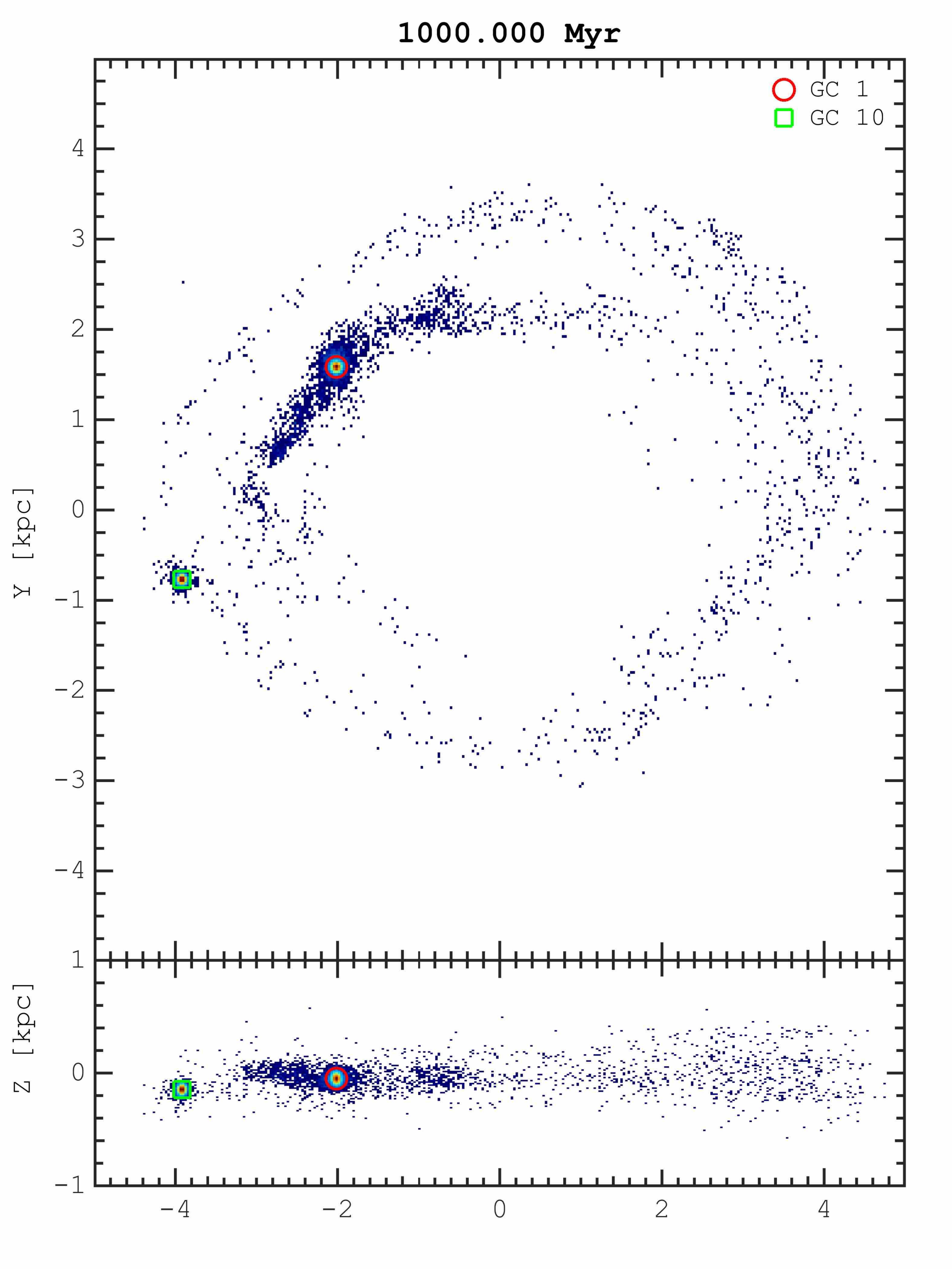}}{mov_swf/dynamics_1_10.mp4.swf}
 \mediabutton[mediacommand=test:playPause]{\fbox{Play/Pause}}
\caption{As in Fig.~5, but for GCs 1 and 10.}\label{fig::app::1-10}
 \end{center}
\end{figure*}

The evolution of some of the globular clusters pairs participated in a single minor mass capture at different times is shown in Figs.~\ref{fig::sm91-100} and \ref{fig::app::1-10}.

\end{appendix}

\end{document}